\documentclass{iaus}
\usepackage{graphicx}

\title[IAU 275.~~Blazar jet physics in the age of Fermi] 
{Blazar jet physics in the age of Fermi}

\author[Charles D.\ Dermer]   
{Charles D.\ Dermer$^1$, on behalf of the Fermi LAT Collaboration}

\affiliation{$^1$Naval Research Laboratory, Code 7653,  \\ 
4555 Overlook Ave., SW, Washington, DC 20375-5352 USA \\ 
email: {\tt charles.dermer@nrl.navy.mil} }

\pubyear{2010}
\volume{275}  
\pagerange{xxx--yyy}
\jname{Jets at all Scales}
\editors{G.E.\ Romero, R.A.\ Sunyaev and T.\ Belloni, eds.}
\begin{document}

\maketitle

\begin{abstract}

The impact of the Fermi Gamma-ray Space Telescope on blazar research is reviewed. This includes a brief description of the Fermi Large Area Telescope, a summary of the various classes of extragalactic sources found in the First Large Area Telescope AGN Catalog, and  more detailed discussion of the flat spectrum radio quasar 3C454.3 and the BL Lac object PKS 2155-304. Some theoretical studies related to ongoing blazar research with Fermi are mentioned, including implications of $\gamma$-ray observations of radio galaxies on blazar unification scenarios, variability in colliding shells, and whether blazars are sources of ultra-high energy cosmic rays.

\end{abstract}

\firstsection 
\section{Introduction}

Even without the Fermi Gamma-ray Space Telescope, the study of blazar jets would be rapidly advancing from discoveries made with ground-based $\gamma$-ray Cherenkov telescopes and AGILE. With Fermi, we have a telescope that is giving the most complete census yet of the $\gtrsim 100$ MeV sky. Fermi makes sensitive measurements of the $\gamma$-ray spectra of blazars where the energy output often dominates. Unexpected puzzles have arisen. This includes the origin of the multi-GeV cutoffs in 3C 454.3 and other FSRQs. Another is the relation of blazars to radio galaxies detected at $\gamma$-ray energies. The blazar sequence (\cite[Fossati et al.\ 1998)]{fos98}, if it is that, remains sketchily understood, even with our new knowledge from Fermi.

Here we briefly review the results of blazar studies in the age of Fermi, including the 1$^{\rm st}$ Fermi LAT AGN Catalog (1LAC; \cite[Abdo et al.\ 2010a]{abd10a})  and the misaligned AGN (MAGN) sample of radio galaxies detected with Fermi. An overview of various classes of galaxies and AGNs detected with Fermi is given, including some detail on individual sources. 3C 454.3 and PKS 2155-304 Theoretical questions arising from these observations are raised, and some speculations in terms of colliding shells ejected from the central supermassive black hole are made. The likelihood that blazars accelerate ultra-high energy cosmic rays is considered.

\section{The Large Area Telescope on Fermi}

The Large Area Telescope (LAT) on the Fermi Gamma-ray Space Telescope is a pair conversion telescope that tracks the direction of an electron-positron pair formed by an incident $\gamma$ ray. The converter-tracker is made of a 4$\times$4 array of 16 modules. Each module has 18 tracker planes that uses Si strips to follow the direction of the converted pair. The upper 16 tracker planes have tungsten converters, with the uppermost 12 ``front" planes using thin layers of tungsten to minimize multiple scattering and provide a narrow point spread function (psf), and the lower 4 ``back" planes using thick tungsten layers to increase the effective area at the expense of the psf. The bottom two planes have no converters, only Si tracking planes. The energy is deposited in the calorimeter, which is made of 96 CsI(Tl) crystals. The entire assembly is surrounded by a segmented anti-coincidence detector used to reject charged particles. See \cite[Atwood et al.\ (2009)]{atw09} for more detail.

The LAT has a field-of-view, roughly defined by the solid angle within which the effective area is more than one-half of the on-axis effective area, of 2.4 sr, which corresponds to $\approx 1/5^{\rm th}$ of the full sky. For the most stringent {\it diffuse} event class used to identify $\gamma$ rays, the on-axis effective area increases from $\approx 1500$ cm$^2$ at 100 MeV to $\approx 8000$ cm$^2$ at 1 GeV, and is roughly constant at higher energies. The energy resolution is better than 10\% between $\approx 50$ MeV and $\approx 50$ GeV. For front-converting events, the 68\% acceptance cone is $\approx 3.5^\circ$ and $\approx 0.6^\circ$ at photon energies $E\approx 100$ MeV and $E\approx 1$ GeV, respectively.  Updated instrument performance that improves on pre-flight Monte Carlo and muon calibrations by using inflight corrections to the instrument response functions are found at http://www-glast.slac.stanford.edu/software/IS/glast$\_$lat$\_$performance.htm. 

Fermi nominally operates in a zenith-pointing rocking mode that allows the LAT to survey the entire sky every two orbits, or $\approx$ every 3 hours.  It typically rocks 35$^\circ$ north and south of the zenith on alternate orbits. Autonomous repoints to gamma-ray bursts and two dedicated pointings, one to 3C 454.3 and a second to the flaring Crab, have been executed by Fermi.

\section{LBAS and 1LAC}

Two lists of AGNs detected with Fermi have now been published by the Fermi Collaboration. After 3 months of science observations between 2008 August 4 and 2008 October 30, the LAT Bright AGN Sample (LBAS) was released (\cite[Abdo et al.\ 2009a]{abd09a}). It consists of 106 high-latitude ($|b|>10^\circ$) sources associated with AGNs. These sources have a test statistic $TS>100$, corresponding to $\gtrsim 10\sigma$ significance, and are a subset of the 205 sources listed in the Fermi LAT bright source list (\cite[Abdo et al. 2009b]{abd09b}). By comparison, the 3$^{\rm rd}$ Egret Catalog (3EG; \cite[Hartman et al.\ 1999]{har99}) of $\gamma$-ray sources and the revised EGRET catalog (EGR; \cite[Casandjian \& Grenier 2008]{cg08}) list 31 sources with significance $>10\sigma$, of which 10 are at high latitude. Remarkably, 5 of the $>10\sigma$ EGRET sources are not found in the LAT bright source list. These are the flaring blazars NRAO 190, NRAO 530, 1611+343, 1406-076, and PKS 1622-297,  the most luminous EGRET blazar (\cite[Mattox et al.\ 1997)]{mat97}.

The 1LAC (\cite[Abdo et al.\ 2010a]{abd10a}) is a subset of the 1451 sources in the First LAT Fermi Source Catalog (1FGL; \cite[Abdo et al.\ 2010b]{abd10b}) derived from analysis of data taken during 11 months of observation between 2008 August 4 and 2009 July 4. There are 1043 1FGL sources at high latitudes, of which 671 are associated with 709 AGNs, with the larger number of AGNs than 1LAC sources due to multiple associations. Associations are made by comparing the localization contours with counterparts in various source catalogs, for example, the flat-spectrum 8.4 GHz CRATES (Combined Radio All-Sky Targeted Eight GHz Survey; \cite[Healey et al.\ 2007]{hea07}) and the Roma BZCAT blazar catalog (\cite[Massaro et al.\ 2009]{mas09}). The probability of association is calculated by comparing the likelihood of chance associations with catalog sources if randomly distributed. Positional coincidence only affords an association; correlated flux variability between different wavebands is required for a firm identification.

\begin{table}
  \begin{center}
  \caption{Classes of $\gamma$-ray emitting AGNs and galaxies in the 1LAC ``clean" sample}
  \label{tab1}
 {\scriptsize
  \begin{tabular}{|l|c|c|c|c|}\hline 
{\bf Class} & {\bf Number} & {\bf Characteristics} & {\bf Prominent Members} & {\bf Other} \\  \hline
All & 599 &  &  &  \\ \hline
BL Lac objects & 275 & weak emission lines &  AO 0235+164 &  \\
 \dots LSP  & 64 & $\nu^{\rm syn}_{pk}< 10^{14}$ Hz & BL Lacertae & \\ 
 \dots ISP  & 44 & $10^{14}$ Hz $<\nu^{\rm syn}_{pk}< 10^{15}$ Hz& 3C 66A, W Comae & \\
 \dots HSP  & 114 & $\nu^{\rm syn}_{pk}> 10^{15}$ Hz &  PKS 2155-304, Mrk 501 & \\ \hline
FSRQs & 248 & strong emission lines & 3C 279, 3C 354.3  &  \\
 \dots LSP  & 171 & & PKS 1510-089& \\ 
 \dots ISP  & 1 & & & \\
 \dots HSP  & 1 &  & & \\ \hline
New Classes$^1$ & 26 &  &  &  \\
 \dots Starburst  & 3 & active star formation & M82, NGC 253 & \\ 
 \dots MAGN  & 7 & steep radio spectrum AGNs & M87, Cen A, NGC 6251 & \\
 \dots RL-NLS1s  & 4 & strong FeII, narrow permitted lines & PMN J0948+0022  & \\
 \dots NLRGs  & 4 & narrow line radio galaxy & 4C+15.05& \\ 
 \dots other sources$^2$  & 9 & & & \\ \hline
Unknown & 50 &  &    &  \\ \hline
  \end{tabular}
  }
 \end{center}
\vspace{1mm}
 \scriptsize{
$^1$Total adds to 27, because the RL-NLS1 source PMN J0948+0022 is also classified as an FSRQ in the 1LAC\\
$^2$Includes PKS 0336-177, BZU J0645+6024, B3 0920+416, CRATES J1203+6031, CRATES J1640+1144, 
CGRaBS J1647+4950, B2 1722+40, 3C 407, and 4C +04.77}
\end{table}

Of the 671 associations, 663 are considered ``high-confidence" associations due to more secure positional coincidences. The ``clean" sample is a subset of the high-confidence associations consisting of 599 AGNs with no multiple associations or other peculiarities.  As listed in Table 1, these subdivide into 275 BL Lac objects, 248 flat spectrum radio quasars, 26 other AGNs, and 50 AGNs of unknown types. The ``New Classes"  category contains non-blazar AGNs, including starburst galaxies and radio galaxies of various types (e.g., narrow line and broad line). An AGN is classified as an ``unknown" type either because it lacks an optical spectrum, or the optical spectrum has insufficient statistics to determine if it is a BL Lac objects or a FSRQ. In comparison with the 671 AGNs in the 1LAC taken with 11 months of Fermi data, EGRET found 66 high-confidence ($>5\sigma$) detections of blazars out of total of 271 sources in the 3EG, with another 27 lower-confidence detections with significance between $4\sigma$ and $5\sigma$. Thus the 1LAC already represents an order-of-magnitude increase in the number of AGNs over EGRET. There are $\approx 300$ -- 400 unassociated and therefore unidentified high-latitude Fermi sources in the LBAS.

\section{Classification of radio-emitting AGNs and unification}

Different classes of extragalactic AGNs are defined according to observing frequency. We already noted the association of Fermi sources with BL Lac objects and flat spectrum radio quasars (FSRQs). This represents an optical classification.  The precise definition used by the Fermi team is that an AGN is a BL Lac object if the equivalent width of the strongest optical emission line is $<5$\AA, and the optical spectrum shows a Ca II H/K break ratio  $< 0.4$ in order to ensure that the radiation is predominantly nonthermal (the Ca II break arises from old stars in elliptical galaxies).  The wavelength coverage of the spectrum must satisfy $(\lambda_{max} - \lambda_{min})/\lambda_{max} > 1.7$ in order that at least one strong emission line would have been detected if present. This helps guard against biasing the classification for AGNs at different redshifts where the emission lines could be redshifted out of the relevant wavelength range. For sources exhibiting BL Lac or FSRQ characteristics at different times, the criterion adopted is that if the optical spectrum conforms to BL Lac properties at any time, then it is classified as a BL Lac object.  

The criterion for classification of radio galaxies according to their radio properties stems from the remarkable correlation between radio morphology and radio luminosity (\cite[Fanaroff \& Riley 1974)]{fr74}. The twin-jet morphology of radio galaxies is seen in low-power radio galaxies, whereas the lobe and edge-brightened morphology is found in high-power radio galaxies, with a dividing line at $\approx 2\times 10^{25}$ W/Hz at 178 MHz, or at a bolometric radio luminosity of $\approx 2\times 10^{40}$ erg s$^{-1}$. Besides a radio-morphology/radio-power classification, radio spectral hardness can also be used to characterize sources as flat-spectrum and steep-spectrum sources.  The misaligned AGNs (MAGNs) are 1LAC sources associated with steep ($F_\nu \propto \nu^{-\alpha_r}$, with $\alpha_r > 0.5$) radio-spectrum objects typically at 178 MHz in the Third Cambridge and Molonglo radio catalogs that show extended radio structures in radio maps. By contrast, some radio galaxies show hard radio spectra in the 1 -- 10 GeV range, and can be subdivided according to the widths of the optical emission lines into broad- and narrow-line radio galaxies (\cite[Perlman et al.\ 1998]{per98}). Because of the close relation between core dominance and $\gamma$-ray spectral properties, core dominance can be used to infer the alignment of a radio-emitting AGN  (e.g., \cite[Lister \& Homan 2005)]{lh05}.
 
Blazars and radio galaxies can also be classified according to their broadband spectral energy distribution (SED) when there is sufficient multiwavelength coverage to reconstruct a spectrum from the radio through the optical and X-ray bands. When the peak frequency $\nu^{\rm syn}_{pk}$ of the synchrotron component of the spectrum is $< 10^{14}$ Hz, then a source is called low synchrotron-peaked (LSP), whereas if the SED has $\nu^{\rm syn}_{pk}> 10^{15}$ Hz, then it is referred to as high synchrotron-peaked (HSP). Intermediate synchrotron-peaked (ISP) objects have $10^{14}$ Hz $<\nu^{\rm syn}_{pk}< 10^{15}$ Hz. SEDs of the bright Fermi LBAS sources are constructed in \cite[Abdo et al.\ (2010c)]{abd10c}. Essentially all FSRQs are LSP blazars, whereas BL Lac objects sample the LSP, ISP, and HSP range.

According to the standard unification scenario for radio-loud AGNs (\cite[Urry 
\& Padovani 1995)]{up97}, radio galaxies are misaligned blazars, and FR1 and FR2 radio galaxies are the parent populations of BL Lac objects and FSRQs, respectively. To establish this relationship requires a census of the various classes of sources that takes into account the different beaming properties for the Doppler-boosted radiation of blazars. Even if analysis of data of radio galaxies and blazars supports the unification hypothesis, this paradigm still does not explain the reasons for the differences between radio-quiet and radio-loud AGNs, or between BL Lac objects and FSRQs.

Other classes of extragalactic Fermi sources found in the 1LAC include starburst galaxies, narrow line radio galaxies (NLRGs), radio-loud narrow line Sy 1s (RL-NLS1s), and radio-quiet AGNs.  The recent $\gamma$-ray detections of the starburst galaxies M82, NGC 253, and NGC 4945 (though the latter source has a Sy 2 nucleus that might contribute to the $\gamma$-ray emission) with Fermi, VERITAS, and HESS opens up a new avenue of research in cosmic-ray studies. Five NLRGs are reported in the 1LAC. These objects have narrow emission lines in their optical spectrum, suggesting that they are observed at large angles with respect to the jet direction, with the surrounding dust torus obscuring the broad line region (BLR). 

RL-NLS1s have also been recently established as a $\gamma$-ray source class (\cite[Abdo et al.\ 2009c]{abd09c}). These objects show narrow H$\beta$ lines with FWHM line widths $\lesssim 2000$ km s$^{-1}$, weak forbidden lines ($[OIII]/H\beta < 3$) 
and a strong Fe II bump, and are therefore classified as narrow-line type I Seyferts (\cite[Pogge 2000)]{pog00}. By comparison with the $\sim 10^9 M_\odot$ black holes in blazars, the host galaxies of RL-NLS1s are spirals that host nuclear black holes with relatively small ($\sim 10^6$ -- $10^8 M_\odot$) mass that accrete at a high Eddington rate. The detection of these objects challenges scenarios where radio-loud AGNs are hosted by elliptical galaxies that form as a consequence of galaxy mergers.

The 1LAC includes 10 associations with radio-quiet
AGNs. Radio-quiet AGNs have not yet been established 
as a $\gamma$-ray source class. In 8 of these cases, at least one 
blazar, radio galaxy, or CRATES source is also found close
to the $\gamma$-ray source. In the remaining two cases, the
association probabilities are weak. Thus none appear in 
the 1LAC ``clean" sample. It remains unclear whether 
any radio-quiet sources, including Sy 2 galaxies such as 
NGC 4945 or NGC 1068 (\cite[Lenain et al.\ 2010]{len10}), produce
$\gamma$ rays from Sy nuclei rather than cosmic-ray processes.

\section{Properties of Fermi AGNs}

Various correlations are found by comparing $\gamma$-ray properties of Fermi AGNs according to their radio, optical, or SED classification. Probably the most pronounced correlation is between the $\gamma$-ray spectral index $ \Gamma_\gamma $ measured between 100 MeV and 100 GeV, and optical AGN type. FSRQs have significantly softer spectra than BL Lac objects, with $\langle \Gamma_\gamma \rangle \cong 2.40\pm 0.17$ for FSRQs and $\langle \Gamma_\gamma \rangle \cong 1.99\pm 0.22$ for BL Lac objects in the LBAS (\cite[Abdo et al.\ 2009a]{abd09a}). The SED classification shows that the $\gamma$-ray spectral index progressively hardens from $\langle \Gamma_\gamma \rangle \cong 2.48, 2.28, 2.13,$ and 1.96 when the class varies from FSRQs to LSP-BL Lacs, ISP-BL Lacs, and HSP-BL Lacs, respectively.

Fermi data reveal complex $\gamma$-ray blazar spectra. All FSRQs and LSP-BL Lac objects, and most ISP blazars show breaks in the $\approx 1$ -- 10 GeV range (\cite[Abdo et al.\ 2010d)]{abd10d}. Such a break was apparent from the first observation of the bright blazar 3C 454.3 (\cite[Abdo et al.\ 2009d)]{abd09d}, and will be discussed in more detail below. This has the unfortunate consequence, however, to reduce the utility of the FSRQs for EBL studies. The HSP blazars, though, are generally well-described by a flat or rising $\nu F_\nu$ SED in the GeV range, with spectral breaks between $\approx 10$ -- 100 GeV implied from $\gtrsim 100$ GeV measurements with air Cherenkov telescopes. 

There is a much higher ratio, close to unity, of BL Lacs to FSRQs in the 1LAC compared to the 3EG, where the ratio was $\approx 25$\%. This is in large part due to the better multi-GeV sensitivity of Fermi over EGRET which finds far more HSP BL Lac objects than with EGRET, though stronger redshifting effects on the higher-redshift FSRQs than BL Lacs could also contribute. Only 121 out of 291 BL Lac objects had measured redshifts at the time of publication of the 1LAC.  For sources with measured redshift $z$, BL Lac objects are mostly found at low ($z\lesssim 0.4$) redshifts, with only a few HSP BL Lac objects at higher redshifts. By contrast, the FSRQs span a wide range from $z\approx 0.2$ to the highest redshift 1LAC blazar with $z = 3.10$.

\begin{figure}[t]
\begin{center}
 \includegraphics[width=3.4in]{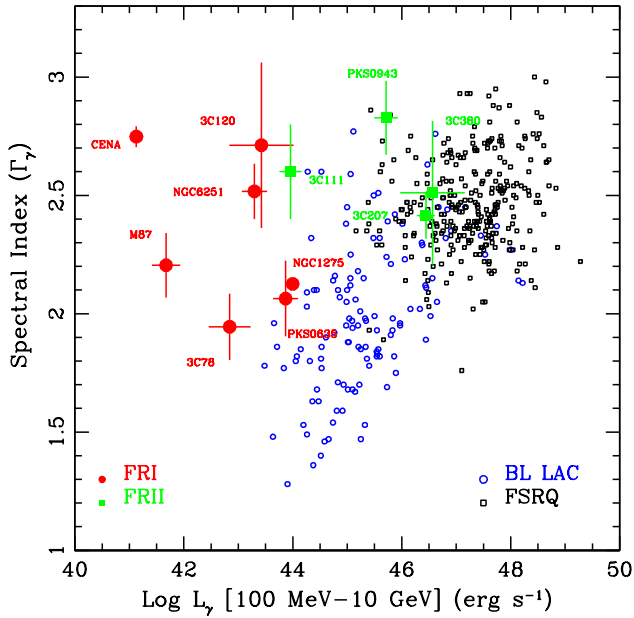} 
 \caption{Gamma-ray spectral slopes of FR1 radio galaxies (red circles), FR2 radio sources (green squares), BL Lac objects (open blue circles) and FSRQs (open black squares), are plotted as a function of their 100 MeV -– 10 GeV $\gamma$-ray luminosity (\cite[Abdo et al.\ 2010e]{abd10e}). The $\gamma$-ray emitting misaligned AGNs (MAGNs) are the red and green points.}
\label{fig1}
\end{center}
\end{figure}

This significant redshift incompleteness hampers interpretation of AGN properties, and the overall behavior and explanation of the blazar sequence (see discussion in Abdo et al.\ 2010a), and the role of RL NLS1s in understanding the sequence \cite[Foschini et al.(2009)]{fos09}, if it is not 
due to selection biases (\cite[Padovani et al.\ 2003)]{pad03}). Nevertheless, a plot of $\Gamma_\gamma $ as a function of apparent isotropic $\gamma$-ray luminosity $L_\gamma$ in the 100 MeV -- 10 GeV band for those sources with measured redshift can be constructed (Fig.\ 1; from \cite[Abdo et al.\ 2010e]{abd10e}). It shows that the hard-spectrum BL Lac objects typically have much lower $L_\gamma$ than the FSRQs. This  divide has been interpreted as a change in the accretion regime at approximately 1\% of the Eddington luminosity (\cite[Ghisellini et al.\ 2009]{ghi09}). In addition, the nearby radio galaxies with $z\lesssim 0.1$ inhabit a separate portion of the $\Gamma_\gamma$ vs.\ $L_\gamma$ plane, and are characterized by lower $\gamma$-ray luminosities than their parent populations. The values of $L_\gamma$ and $\Gamma_\gamma$ of the two more distant steep spectrum radio sources, 3C 207 ($z = 0.681$) and 3C 380 ($z = 0.692$), and the FR2 radio galaxy PKS 0943-76 ($z = 0.27$) fall, however, within the range of $\gamma$-ray luminosities measured from FSRQs. 

Both core and lobes can significantly contribute to the measured $\gamma$-ray luminosities. In the case of Centaurus A, the values of $L_\gamma$ of the core and lobes are comparable, with the lobe emission primarily attributed to Compton-scattered CMBR
(\cite[Abdo et al.\ 2010f)]{2010Sci...328..725F}. The significant or dominant lobe component means that the core luminosity of misaligned AGNs can be less than the measured $L_\gamma$ unless the lobe emission is resolved.

\subsection{3C 454.3}

The FSRQ 3C 454.3, at $z = 0.859$, underwent giant flares and became the brightest $\gamma$-ray source in the sky for a week in 2009 December and again in 2010 April  (\cite[Ackermann et al.\ 2010)]{ack10}. The latter outburst triggered a pointed-mode observation by Fermi. During the December outburst, its daily flux reached $F= 22 (\pm 1)\times 10^{-6}$ ph($> 100$ MeV) cm$^{-2}$  s$^{-1}$, corresponding to an apparent isotropic luminosity of $\approx 3\times 10^{49}$ erg s$^{-1}$, making it the most luminous blazar  yet detected with Fermi.  Using the measured flux and a one-day variability timescale at the time that the most energetic photon (with energy $E\approx 20$ GeV) was detected, implies a minimum Doppler factor of $\delta_{\rm D,min}\approx 13$. Assuming that the outflow Lorentz factor $\Gamma \approx 20$, consistent with the inferred value of $\delta_{\rm D,min}$ and with radio observations at a different epoch (\cite[Jorstad et al.\ 2005]{jor05}), then simple arguments suggests a location $R \lesssim c \Gamma^2 t_{\rm var}/(1+z)\approx 0.2 (\Gamma/20)^2 (t_{\rm var}/{\rm day})$ pc, which is at the outer boundary of the BLR. Flux variability on time scales as short as 3 hr was measured at another bright flux state, which suggests that the $\gamma$-ray emission site would be even deeper in the BLR. This stands in contrast to inferences based on coherent optical polarization changes over timescales of weeks prior to a $\gamma$-ray flare in 3C 279  (\cite[Abdo et al.\ 2010g)]{abd10g}, which seems to place the emission site at much larger distances. 

\begin{figure}[t]
\begin{center}
\includegraphics[width=5.0in,height=3.0in]{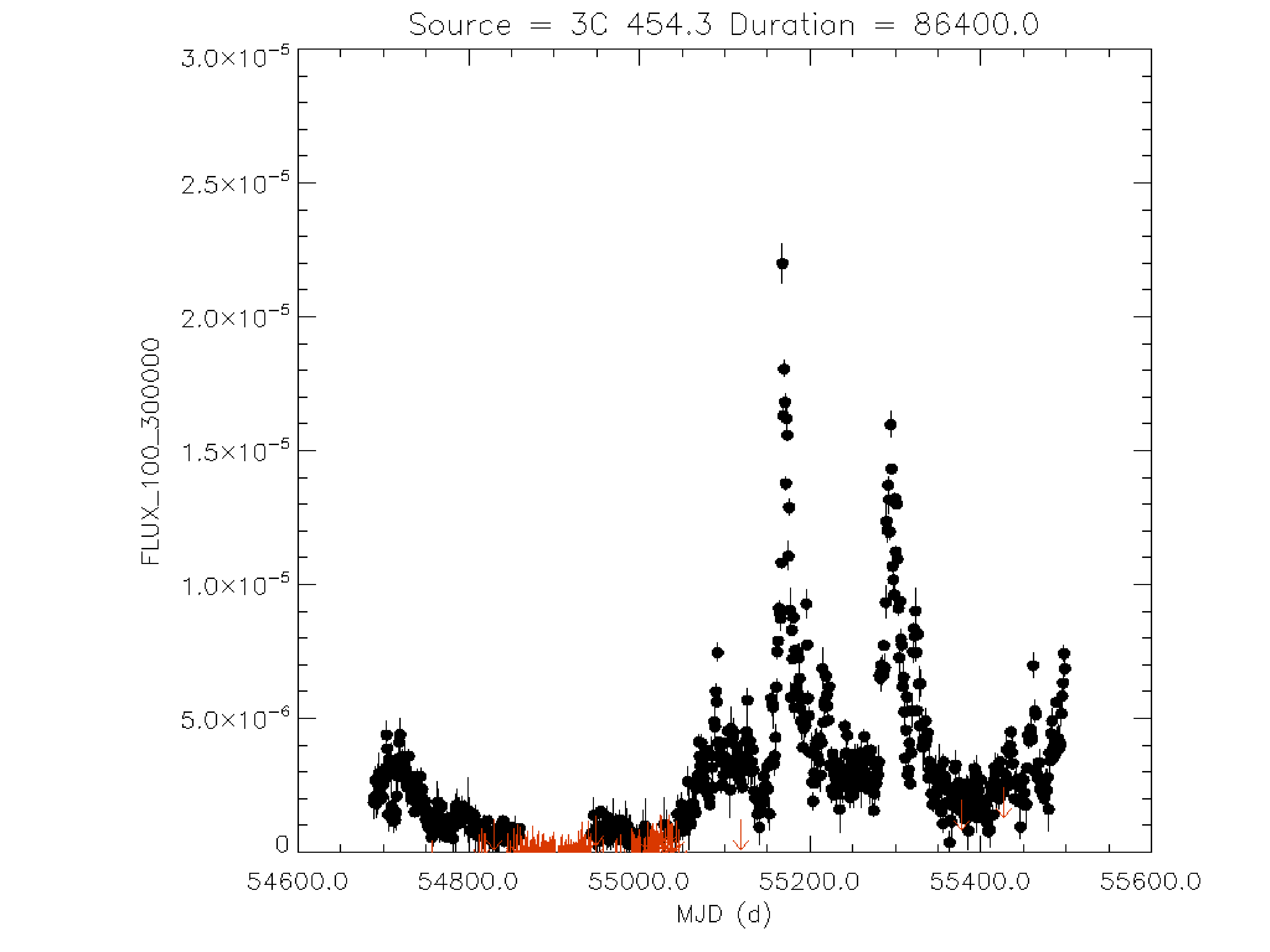} 
 \caption{Fermi LAT 1-day light curve of 3C 454.3, showing the giant flares in December 2009, and April 2010  (MJD 55200 corresponds to 4 January 2010). }
\label{fig1}
\end{center}
\end{figure}

Figure 2 shows the light curve of 3C 454.3 taken from the public website on Fermi monitored sources\footnote{fermi.gsfc.nasa.gov/ssc/data/access/lat/msl$\_$lc/}, measured in durations of one day over the course of the Fermi mission.  Intense flaring occurs during periods of enhanced activity, as if the engine is being fueled prior to reaching a state coinciding with maximum energy output. Indeed, $\gamma$-ray flux enhancements reach a plateau preceding a major flare, and the 2008 July outburst shows strong resemblance to those in 2009 August and  2010 December. 

As noted already in the initial Fermi report (\cite[Abdo et al.\ 2009d)]{abd09d}, the spectrum of 3C 454.3 breaks strongly by $1.2(\pm 0.3)$ units at $E_{br}\approx 2$ GeV. Such a break is inconsistent with simple radiative cooling scenarios, which predict a break by 0.5 units. The more recent analysis of 3C 454.3 data (\cite[Ackermann et al.\ 2010)]{ack10}, including the major outbursts, confirms the strong spectral break and finds that  $E_{br}$ is very weakly dependent on the flux state, even when the flux changes by more than an order of magnitude. No obvious pattern is found in the spectral index/flux plane, as might be expected in simple acceleration and cooling scenarios (\cite[e.g., Kirk et al.\ 1998)]{krm98}. 

The origin of the spectral break in 3C 454.3 bears on several important issues in FSRQs: the location of the $\gamma$-ray emission site; the source of soft target photons in Compton-scattering models; and the relation of FSRQs and BL Lacs in view of the disappearance of such breaks in ISP and HSP blazars. Such a break would be readily understood if the target field was sufficiently intense to attenuate the blazar radiation by $\gamma\gamma$ absorption processes, but the intense line Ly$\alpha$ radiation field at 10.2 eV observed in GALEX measurements of 3C 454.3 (\cite[Bonnoli et al.\ 2010)]{bon10}  implies $E_{br}\gtrsim 30$ GeV (\cite[Reimer 2007)]{rei07}. Photon attenuation deep within the BLR by He II recombination and Ly$\alpha$ radiation with $E> 54.4$ eV  has been proposed (\cite[Poutanen \& Stern 2010)]{ps10}, but the model lacks a consistent treatment of synchrotron and Compton-scattered blazar flare emission. A full spectral model for the SED of 3C 454.3 can fit the break with a distribution of nonthermal electrons that scatters accretion-disk and BLR radiation (\cite[Finke \& Dermer 2010)]{fd10}, but a solution insensitive to changes in the location of the emission site requires a BLR with a wind-like density. 
 
The spectral break could also be due to Klein-Nishina effects in scattering, 
as has been proposed to explain the SED of PKS 1510-089 (\cite[Abdo et al.\ 2010h)]{abd10h}. The KN break due to upscattered Ly$\alpha$ radiation occurs at a few GeV, and the observed break energy is insensitive to the Doppler factor. But the break is not sharp enough to fit the spectrum of 3C 454.3 spectrum for a power-law electron distribution, and the electron spectrum resulting from cooling is likely to exhibit a hardening due to competition between synchrotron and Compton losses (\cite[Dermer \& Atoyan 2002,]{da02}\cite[ Moderski et al.\ 2005)]{mod05}.

\subsection{PKS 2155-304, HSP BL Lac objects, and FR1 radio galaxies}

PKS 2155-304, an X-ray selected BL Lac object at $z = 0.116$, is one of the most prominent representatives of the HSP blazar population. Its two-peaked SED 
was measured for 11 days between 2008 August 25 and 2008 September 6 at
optical (ATOM), X-ray (RXTE and Swift), and $\gamma$-ray (Fermi and HESS) frequencies. 
This low-state SED is well fit by a one-zone synchrotron/SSC model with  Doppler factor $\delta_{\rm D} = 32$,  magnetic field $B^\prime = 0.018$ G, and comoving 
size $R^\prime = 1.5\times 10^{17}$ cm (corresponding to a variability time of 2 d; \cite[Aharonian et al.\ 2009]{aha09}). During a period of extraordinary flaring on 2006 July 28, PKS 2155-304 exhibited a sequence of $\gamma$-ray flares varying on time scales as short as $\approx 5$ min (\cite[Aharonian et al.\ 2007)]{aha07}. Detailed one-zone synchrotron SSC model fits using a low EBL require $\delta_{\rm D} \gtrsim 100$ in order to avoid  attenuation of the TeV $\gamma$ rays and  provide a good fit to the broadband SED (\cite[Finke et al.\ 2008)]{fin08}.

One-zone synchrotron/SSC models with $\delta_{\rm D} \gtrsim 10$ give good fits to other HSP BL Lac objects such as Mrk 421 and Mrk 501. The SEDs of radio galaxies, in contrast, are fit with much smaller Doppler factors. The SED of the core of Cen A, for instance, can be fit with $\delta_{\rm D} \approx 1$ and bulk Lorentz factors $\Gamma \approx $ few (\cite[Abdo et al.\ 2010i)]{abd10i}. Likewise, the SEDs of the FR1 radio galaxies NGC 1275 (\cite[Abdo et al.\ 2009e]{abd09e}) and M87 (\cite[Abdo et al.\ 2009f]{abd09f}) are well fit with $\delta_{\rm D}\approx 2$ and $\Gamma \sim 4$.   
 
The much larger values of $\Gamma$ for BL Lac objects than for their putative parent population, the misaligned FR1 radio galaxies, is contrary to simple unification expectations. Moreover, even though the $\gamma$-ray luminosities from FR1 radio galaxies are much smaller than that of BL Lac objects (Fig.\ 1), they are still larger than expected by debeaming the radiation of BL Lac objects with $\Gamma \gtrsim 20$.  Additional soft target photons that can be Compton scattered to high energies result in a reduction of the  value of $\delta_{\rm D}$ compared to those implied by the one-zone synchrotron/SSC model. These target photons can  be produced in a structured jet, as in the spine and sheath model (\cite[Chiaberge et al.\ 2000)]{chi00}. Another soft photon source arises if blazar flows decelerate from the inner jet to the pc scale  (\cite[Georganopoulos \& Kazanas 2003]{gk03}), in accord with the mildly relativistic flows at the sub-pc scale found in radio observations of Mrk 421 and Mrk 501.

Given the results of synchrotron/SSC modeling, the Fermi observations suggest that the $\gamma$-ray emission from the core of a radio galaxy originates from a slower region than the emission of BL Lac objects. This can be understood in a colliding shell model if shells with large opening angles $\theta_j$ have lower $\Gamma$ and are less energetic than shells with narrow opening angles. Such a circumstance might also explain the short variability timescale of  PKS 2155-304, as we now show.

Consider a simple colliding shell model where the second shell is much more powerful and has a much larger Lorentz factor than the first shell. Denote the Lorentz factors of the first and second shells by $\Gamma_{a(b)} = 1/\sqrt{1-\beta_{a(b)}^2}$, with $\zeta \equiv \Gamma_b/\Gamma_a \gg 1$. The collision radius takes place at $r_{\rm coll} \approx 2\Gamma_a^2 \Delta t_*$, where $\Delta t_* \gtrsim R_{\rm S}/c$, and ejection time scales are required to be greater than $R_{\rm S}$/c, where $R_{\rm S}$ is the Schwarzschild radius. If the second shell is much more powerful than the first, then a strong forward shock is formed that travels through shell $a$ with Lorentz factor $\bar\Gamma_f$ given by the relative Lorentz factor $\Gamma_{rel}$ of the two shells, so that
$\bar\Gamma_f \cong \Gamma_{rel} = \Gamma_a\Gamma_b(1-\beta_a\beta_b) \rightarrow (\zeta + \zeta^{-1})/2 \rightarrow \zeta/ 2$
(see \cite [Sari \& Piran 1995)]{sp95}.

For short timescale variability, both the radial and angular timescales must be much shorter than $R_{\rm S}/c$.
The radial timescale 
\begin{equation}
t_{rad} = {1+z\over \delta_{\rm D}}\;{\Gamma_a \Delta t_*\over \bar\beta_f \bar\Gamma_f}
\approx {\Gamma_a \Delta t_*\over \Gamma\bar\Gamma_f}\approx 2 \Delta t_*/\zeta^2\;
\end{equation} 
for low-redshift sources, noting that $\Gamma_a \Delta t_*/\bar\Gamma_f$ is the width of shell $a$  in the frame of shell $b$, and shell widths $\sim c \Delta t_*$ in the engine frame are required to be $\gtrsim R_{\rm S}/c$. Here the Doppler factor $\delta_{\rm D}\approx \Gamma \approx \Gamma_b$, where $\Gamma$ is the shocked fluid Lorentz factor.  The angular timescale
\begin{equation}
t_{ang} = {(1+z) r_{\rm coll} \over \Gamma^2 c}\
\approx 2{\Gamma_a^2 \Delta t_*\over \Gamma^2}\approx 2 \Delta t_*/\zeta^2\;.
\end{equation} 
In both cases, the variability timescale can be much shorter than $R_{\rm S}/c$ when $\zeta \gg 1$. 

The criterion for a strong forward shock is that the ratio of apparent luminosities $L_{*b}/L_{*a}\gtrsim \zeta^4$. This may seem an extreme requirement if the opening angles  of the slow and fast shells are the same. But if  $\theta_{a(b)} \propto 1/\Gamma_{a(b)}$, $\theta\sim 1/\Gamma$, a condition arising in simulations of relativistic jets (e.g., \cite[Komissarov et al.\ 2009]{kom09}, and references therein), then excessive energy requirements for the fast shell are mitigated.  The maximum radiative efficiency is $\approx \zeta^2 L_{*a}/2$ (\cite[Dermer \& Razzaque 2010]{dr10}), so this system is necessarily very inefficient. 
This suggestion may not only  resolve the short variability measured in PKS 2155-304, but also, as noted above, possibly relieve the Doppler-factor conflict in unification schemes of blazars and radio galaxies. A more detailed treatment is in preparation. 

\section{Blazars and ultra-high energy cosmic rays}

In Fermi acceleration scenarios, two conditions are required to accelerate particles to ultra-high energies $E \gtrsim 10^{20}$ eV.
The first is that the isotropic power must exceed $\approx 10^{46}\Gamma^2/Z^2$ erg s$^{-1}$, where $\Gamma$ is the 
shocked fluid Lorentz factor, and $Ze$ is the charge. The second is that the sources have luminosity density $\gtrsim 
10^{44}$ erg Mpc$^{-3}$ yr$^{-1}$ within the GZK radius to power UHECRs against photohadronic losses on the CMBR.  Consider the FR1 radio galaxy NGC 1275, at $\approx 75$ Mpc. Its $\gamma$-ray luminosity from the 1LAC is $L_\gamma \cong 1.2\times 10^{44}$ erg s$^{-1}$. For the mildly relativistic outflow speeds from synchrotron/SSC models, it has adequate power to accelerate Fe nuclei. Moreover, it is radiating nonthermal $\gamma$-ray power within this volume with a luminosity density $\approx 20\times 10^{45}$ erg Mpc$^{-3}$ yr$^{-1}$. If a small fraction of this power is channeled into UHECRs, then FR1 radio galaxies and BL Lac objects are favored to be the sources of UHECRs, provided that UHECRs are high-Z ions (\cite[Dermer \& Razzaque 2010]{dr10}).

\section{Summary}

The Fermi LAT is providing a uniform spectral and temporal GeV database with much better sensitivity
than either the EGRET or AGILE missions. Besides BL Lac objects and Flat Spectrum Radio Quasars, 
several new classes of $\gamma$-ray galaxies have been established, including radio galaxies, 
radio-loud narrow line Sy 1 galaxies, and star-forming galaxies. The $\gamma$-ray spectral slope of blazars is 
strongly correlated with $\gamma$-ray luminosity and whether they are BL Lac objects or FSRQs, 
though a large fraction of BL Lac objects still lack redshift measurements. FSRQs exhibit spectral cutoffs
between 1 -- 10 GeV, the reason for which is not well understood. Gamma-ray emitting
 FR1 radio galaxies have much lower $\gamma$-ray luminosities than BL Lac objects, but
not as low as expected for one-zone synchrotron/SSC models. Colliding shells with a range of 
collision Lorentz factors might account for the different $\gamma$-ray luminosities of radio galaxies and 
blazars, and the short variability timescale measured in PKS 2155-304. 
BL Lacs and FR1 galaxies have sufficient power and luminosity density to account for the UHECRs.

\acknowledgments

This work is supported by the Office of Naval Research and NASA. I thank L.\ Foschini for comments and T.\ Piran for criticism. The $Fermi$ LAT Collaboration acknowledges support from a number of agencies and institutes for both development and the operation of the LAT as well as scientific data analysis. These include NASA and DOE in the United States, CEA/Irfu and IN2P3/CNRS in France, ASI and INFN in Italy, MEXT, KEK, and JAXA in Japan, and the K.~A.~Wallenberg Foundation, the Swedish Research Council and the National Space Board in Sweden. Additional support from INAF in Italy and CNES in France for science analysis during the operations phase is also gratefully acknowledged.

\begin{discussion}

\discuss{David Meier}{Early in your talk you mentioned 10 possible RQ objects that have been detected, 
but I did not see them on your $\Gamma_\gamma$ vs. luminosity plot. Where do they lie on that plot?}

\discuss{Chuck Dermer}{The association of radio-quiet AGN with $\gamma$-ray sources is very tentative, as we discuss in the 1LAC paper. Moreover, none of them survive in the ``clean" sample shown in the plot. Because the detection of $\gamma$-ray emission from radio-quiet AGNs could also originate from cosmic-ray processes, as in the case of starburst galaxies, and so might require the detection of $\gamma$-ray flux variability to confirm, the Fermi Collaboration is not yet ready to make a definitive statement. The analysis is ongoing.}

\discuss{Tsvi Piran}{In internal shocks the time scale, because of angular spreading, is the variability 
time of the inner engine. I don't see how you can avoid this in your model.}

\discuss{Chuck Dermer} {The collision radius is set by the speed of the slower shell. The angular spreading is set
by the speed of the shocked fluid which, for a strong forward shock, can be much greater than the speed of the 
slower shell. The smaller opening angle of the high speed, large luminosity shell also reduces the angular spreading
time scale. I agree that this process cannot be very efficient, but if $\sim 10$\% of the Eddington luminosity is 
channeled into a narrow opening angle jet $\sim 10^{-2}$ for $\Gamma\sim 100$ during bright flares, only $\sim 1$\% need be converted to yield $\approx 10^{46}$ erg s$^{-1}$ from PKS 2155-304.}

\discuss{Elisabete M. de Gouveia Dal Pino} {Regarding the production of UHECRs by FR1/BL Lac sources, surely 
they have the ``power" to produce them, but how do the CRs escape from these compact sources?}

\discuss{Chuck Dermer} {The escape of UHECRs without suffering photodisintegration in the AGN radiation fields poses an important constraint on the location of UHECR acceleration sites. In FR1 radio galaxies and BL Lac objects, the broad line region is absent, and the accretion-disk radiation is weak. Whether the host galaxy radiation field poses a challenge to this model will require more  detailed calculations.}

\end{discussion}

\end{document}